\documentclass[preprint,aps,prd,amsmath,amssymb]{revtex4}
\pdfoutput=1
\usepackage{graphicx}
\usepackage{color}
\usepackage{epsfig}
\usepackage{dcolumn}
\usepackage{bm}
\usepackage{bbm}
\usepackage{subfigure}
\usepackage{pxfonts}

\def\lsim{\mathrel{\rlap{\lower4pt\hbox{\hskip1pt$\sim$}}
    \raise1pt\hbox{$<$}}}                
\def\gsim{\mathrel{\rlap{\lower4pt\hbox{\hskip1pt$\sim$}}
    \raise1pt\hbox{$>$}}}                

\newcommand{\GeV}{\, \text{GeV}}
\newcommand{\MeV}{\, \text{MeV}}

\newcommand{\gtwo}{I\kern-.1ex I\,}
\newcommand{\beq}{\begin{eqnarray}}
\newcommand{\eeq}{\end{eqnarray}}
\newcommand{\bpm}{\begin{pmatrix}}
\newcommand{\epm}{\end{pmatrix}}
\newcommand{\CKMud}{1 - \dfrac{1}{2}\lambda^2 -\dfrac{1}{8} \lambda^4}
\newcommand{\CKMus}{\lambda}
\newcommand{\CKMub}{A\lambda^3 (\rho - i \eta)}
\newcommand{\CKMcd}{-\lambda + \dfrac{1}{2} A^2 \lambda^5 [1 - 2 (\rho + i \eta)]}
\newcommand{\CKMcs}{1 - \dfrac{1}{2}\lambda^2 - \dfrac{1}{8}\lambda^4 (1 + 4 A^2)}
\newcommand{\CKMcb}{A \lambda^2}
\newcommand{\CKMtd}{A\lambda^3 (1 - \bar{\rho} - i \bar{\eta})}
\newcommand{\CKMts}{- A \lambda^2 + \dfrac{1}{2} A\lambda^4[1 - 2 (\rho + i\eta)]}
\newcommand{\CKMtb}{1 - \dfrac{1}{2}A^2\lambda^2}
\newcommand{\cl}{\, \rm C.L.}
\begin{document}
\title{\Large  \color{red} Minimal Flavor Constraints for Technicolor}
\date{\today}
\author{Hidenori S. Fukano $^{\color{blue}{\varheartsuit}}$}
\email[E-mail: ]{hidenori@cp3.sdu.dk}
\author{Francesco Sannino $^{\color{blue}{\varheartsuit}}$}
\email[E-mail: ]{sannino@cp3.sdu.dk}
\affiliation{$^{\color{blue}{\varheartsuit}}${ CP}$^{ \bf 3}${-Origins}, 
Campusvej 55, DK-5230 Odense M, Denmark.\footnote{{ C}entre of Excellence for { P}article { P}hysics { P}henomenology  and the {Origins} of bright and dark mass.}}
\begin{flushright}
{\it CP$^3$- Orgins: 2009-8}
\end{flushright}
\begin{abstract}

We analyze the constraints on the the vacuum polarization of the standard model gauge bosons from a minimal set of  flavor observables valid for a general class of models of dynamical electroweak symmetry breaking.  We will show that the constraints have a strong impact on the  self-coupling and masses of the lightest spin-one resonances. Our analysis is applicable to any four and higher dimensional extension of the standard model reducing to models of dynamical electroweak symmetry breaking. \end{abstract}

\maketitle

\section{Minimal Models of  (Extended) Technicolor}

Dynamical electroweak symmetry breaking constitutes one of the best motivated extensions of the standard model (SM) of particle interactions. 

Studies of the dynamics of gauge theories featuring fermions transforming according to higher dimensional representations of the new gauge group has led to several phenomenological possibilities \cite{Sannino:2004qp,Dietrich:2005jn,Dietrich:2006cm,Ryttov:2007sr,Christensen:2005cb} such as (Next) Minimal Walking Technicolor (MWT) \cite{Foadi:2007ue} and Ultra Minimal Walking Technicolor (UMT) \cite{Ryttov:2008xe}.  We will collectively refer to them as minimal models of technicolor. In \cite{Belyaev:2008yj} it was launched a coherent program to investigate different signals of minimal models of technicolor at the Large Hadron Collider experiment at CERN. Here, we also investigated in much detail, among other things also the production of the composite Higgs in association with a SM gauge boson  suggested first in \cite{Zerwekh:2005wh}.  An interesting analysis relevant for the LHC phenomenology of low scale technicolor \cite{Eichten:2007sx} has  appeared \cite{Lane:2009ct}. 

Walking dynamics for breaking the electroweak symmetry was introduced in   
\cite{Eichten:1979ah,Holdom:1981rm,Yamawaki:1985zg,Appelquist:1986an}.  It is worth noting that higher dimensional representations have been used earlier in particle physics phenomenology. Time honored examples are grand unified theories. The possibility of unifying the SM gauge interactions within a technicolor framework has been recently addressed within minimal technicolor models in \cite{Gudnason:2006mk}. The discovery \cite{Sannino:2004qp} that theories with fermions transforming according to higher dimensional representations develop an infrared fixed point (IRFP) for an extremely small number of flavors and colors is intriguing. The dynamics of these theories is being investigated using several analytic methods   not only for $SU(N)$  gauge groups \cite{Sannino:2004qp,Dietrich:2006cm,Ryttov:2007cx} but also for $SO(N)$ and $Sp(2N)$ gauge groups \cite{Sannino:2009aw}.  A better knowledge of the gauge dynamics of several nonsupersymmetric gauge theories has been useful to construct explicit UV-complete models able to break the electroweak symmetry dynamically while naturally featuring small contributions to the electroweak precision parameters \cite{Appelquist:1998xf,Kurachi:2006mu,Harada:2004qn,Harada:2005ru,Foadi:2007ue}. These models are economical since they require the introduction of a very small number of underlying elementary fields and can feature a light composite Higgs \cite{Dietrich:2005jn,Dietrich:2006cm,Hong:2004td}.  Recent analyses lend further support to the latter observation \cite{Doff:2009nk,Doff:2008xx,Doff:2009kq}.  The models feature  also explicit dark matter candidates \cite{Ryttov:2008xe,Gudnason:2006yj,Kainulainen:2006wq,Kouvaris:2007iq} and associated interesting phenomenology \cite{Foadi:2008qv,Nardi:2008ix}. 
Moreover, extensions of the SM featuring a new underlying asymptotic free gauge theory are naturally unitary at any arbitrary high energy scale. This strongly increases the theoretical appeal of these extensions. Another important aspect is that the underlying gauge theories can already be tested via first principle lattice computations \cite{Catterall:2007yx,Catterall:2008qk,
Shamir:2008pb,DelDebbio:2008wb,DelDebbio:2008zf, Appelquist:2007hu,Hietanen:2008vc,Hietanen:2008mr,Hietanen:2009az,Deuzeman:2009mh,DelDebbio:2009fd, Fodor:2009wk}. Effective approaches, i.e. four and higher dimensional ones are to be considered as approximations of an underlying dynamics a la Technicolor or of  an unspecified dynamics, see  \cite{Carone:2007md,Hirn:2008tc,Dietrich:2008ni,Nunez:2008wi} for recent efforts.

Whatever is the dynamical extension of the SM it will, in general, modify the vacuum polarizations of the SM gauge bosons. LEP I and II data provided direct constraints on these vacuum polarizations \cite{Peskin:1991sw,Barbieri:2004qk,Buras:1994ec}. In this work we show that we can use flavor physics to provide stronger constraints than previously obtained for some of the precision observables. Our results are in agreement with the analysis made in \cite{Bona:2007vi,Antonelli:2009ws}. 

We are not attempting to provide a full theory of flavor but merely estimate the impact of  a new dynamical sector, per se, on well known flavor observables. We will, however, assume that whatever is the correct mechanism behind the generation of the mass of the SM fermions it will lead to SM type Yukawa interactions \cite{Chivukula:1987py}.  This means that we will constrain models of technicolor with extended technicolor interactions \cite{Dimopoulos:1979es,Eichten:1979ah} entering in the general scheme of minimal flavor violation theories \cite{D'Ambrosio:2002ex}. 
To be specific we will show that it is possible to provide strong constraints on the  technirho and techniaxial self-couplings and masses for a general class of models of dynamical electroweak symmetry breaking. 
Our results can be readily applied to any extension of the SM featuring new heavy spin-one states. In particular it will severely limit the possibility to have very light spin-one resonances to occur at the LHC even if the underlying gauge theory has vanishing $S$-parameter.

\section{Minimal $\Delta F=2$ Flavor Corrections from Technicolor}

Our goal is to compute the minimal contributions, i.e. coming just from the technicolor sector, for processes in which the flavor number $F$ changes by two units, i.e. $\Delta F = 2$. Here we consider  $F$ to be either the strange or the bottom number.  Besides the intrinsic technicolor corrections to flavor processes  one has also the corrections stemming out from extended technicolor models  which are directly responsible for providing mass to the SM fermions.  We will make the assumption, strongly supported by experiments,  that if this extended model exists it leads to a Yukawa sector similar to the SM one. It is, hence, by construction an extended technicolor model implementing the minimal flavor violation  \cite{D'Ambrosio:2002ex} idea. To be more specific we will determine the effects of heavy spin-one resonances mixing with the SM gauge bosons on flavor observables. We use the effective Lagrangian framework presented in \cite{Foadi:2007ue} according to which the relevant interactions  of the composite Higgs sector to the SM quarks up and down reads: 
\beq
{\cal L}^{\rm quark}_{\rm yukawa}
= 
\frac{\sqrt{2}\,\, m_{u_i} }{v} V_{ij} \cdot \bar{u}_{Ri} \pi^+  d_{Lj}
-
\frac{\sqrt{2}\,\, m_{d_i} }{v} V^*_{ji} \cdot \bar{d}_{Ri} \pi^- u_{Lj} 
+ h.c.\,,\label{yukawa-MWT}
\eeq 
where 
$m_{ui} , (u_i=u,c,t)$ and 
$m_{di} , (d_i = d,s,b)$ are respectively the up and down quark masses of the $i^{\rm th}$ generation. 
 $V_{ij}$ is the $i,j$ element of the Cabibbo-Kobayashi-Maskawa (CKM) matrix. 
This is our starting point which will allow us to compute the $\Delta F =2$ processes. We have also checked our results using the Hidden Local Gauge Symmetry \cite{Bando:1987br} version of  \cite{Foadi:2007ue}.  
\begin{figure}[h]
\begin{center}
\begin{tabular}{cc}
{
\begin{minipage}{0.5\textwidth}
\begin{flushleft}
  \hspace*{0.1cm} {(a)}
\end{flushleft}
  \vspace*{-0.5cm}
\hspace*{2ex}
\includegraphics[scale=0.8]{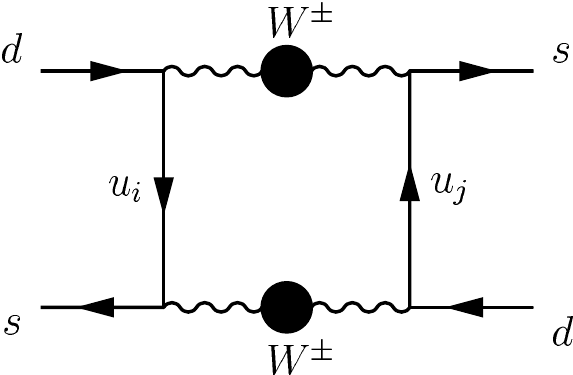}
\end{minipage}
}
{
\begin{minipage}{0.5\textwidth}
\begin{flushleft}
  \hspace*{0.1cm} {(b)}
\end{flushleft}
  \vspace*{-0.5cm}
\hspace*{2ex}
\includegraphics[scale=0.8]{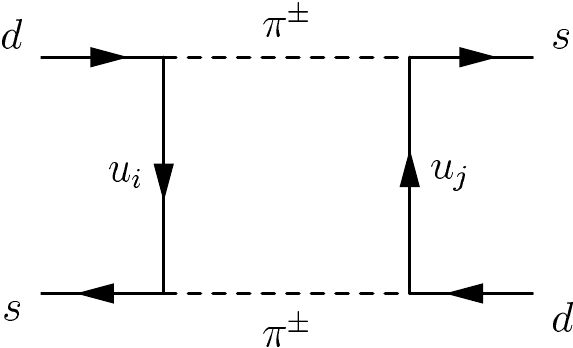}
\end{minipage} 
} \\[2cm]
{
\begin{minipage}{0.5\textwidth}
\begin{flushleft}
  \hspace*{0.1cm} {(c)}
\end{flushleft}
  \vspace*{-0.5cm}
\hspace*{2ex}
\includegraphics[scale=0.8]{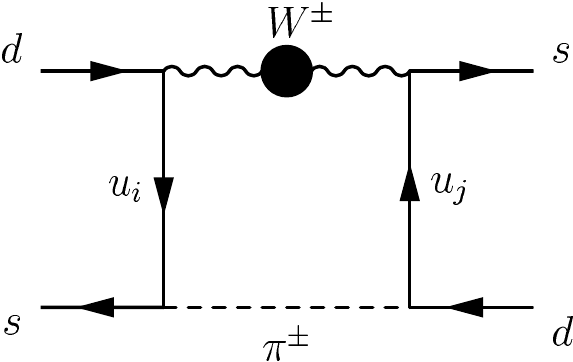}
\end{minipage}
}
{
\begin{minipage}{0.5\textwidth}
\begin{flushleft}
  \hspace*{0.1cm} {(d)}
\end{flushleft}
  \vspace*{-0.5cm}
\hspace*{2ex}
\includegraphics[scale=0.8]{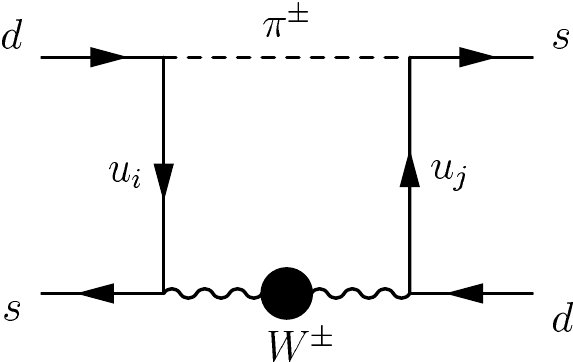}
\end{minipage} 
} 
\end{tabular}
\caption{Box diagrams for $\Delta S = 2$  {\it annihilation} processes. To obtain 
                 the $\Delta B = 2$ process, we should simply rename $s$  with $ b$ and $d$  with $ q\,(q=d,s)$ in the various diagrams.
                 \label{Box-annihilation}}
\end{center}
\end{figure}%

The diagrams contributing to the $\Delta F=2$ process  are shown in Fig.~\ref{Box-annihilation}. They amount to the  {\it annihilation} process \footnote{
Note that the contribution of the { \it scattering} process to the invariant amplitude is equal to that of the {\it annihilation} one.}.

The final contribution to the  $\Delta F =2$ amplitude is: 
\beq
i {\cal M}(\Delta S =2) 
&=& 
2 \times 
\left[ i {\cal M}^{(a)}_\square + i {\cal M}^{(b)}_\square +i {\cal M}^{(c)}_\square +i {\cal M}^{(d)}_\square \right]
\nonumber\\
&=&
2 \times \left( i \, \frac{g_{\rm EW}}{\sqrt{2}}\right)^4  \times 
\frac{-i}{16\pi^2 M^2_W} \times 
\left[ \sum_{i,j=u,c,t} \lambda_i\lambda_j  E(m_i,m_j,M_V,M_A) \right] 
\times Q_{\Delta F=2}\,,
\eeq
where $m_i , (i=u,c,t)$ indicates the $u_i$ mass while 
$M_V,M_A$ are respectively the mass of the lightest techni-vector meson and techni-axial vector one.  
$Q_{\Delta F=2}$ is short for
\beq
Q_{\Delta F=2} =
\begin{cases}
(\bar{s}_L \gamma^\mu d_L)(\bar{s}_L \gamma_\mu d_L) & \text{for $F=S$}\,, \\[2ex]
(\bar{b}_L \gamma^\mu q_L)(\bar{b}_L \gamma_\mu q_L) & \text{for $F=B$}\, .
\end{cases}
\eeq
We introduced the quantity $\lambda_i$:
\beq
\lambda_i =
\begin{cases}
V_{id} V^*_{is} & \text{for $K^0-\bar{K}^0$ system}\,, \\[2ex]
V_{iq} V^*_{ib} & \text{for $B^0_q-\bar{B}^0_q$ system}\,,
\end{cases}
\eeq
encoding the information contained in the CKM matrix.
Moreover, $E(m_i,m_j,M_V,M_A)$  keeps track of the technicolor-modified gauge bosons propagators. Its cumbersome full expression is reported in the technical appendix.  

It is convenient to rewrite the induced $\Delta F=2 $ term of the Lagrangian as follows:
\beq
{\cal L}^{\Delta F=2}_{\rm eff} 
=
- \frac{G^2_F M^2_W}{4 \pi^2} \cdot A(a_V,a_A) \cdot Q_{\Delta F=2}\,,
\label{lag-eff-dF=2}
\eeq
with
\beq
A(a_V,a_A)
\equiv
\sum_{i,j=u,c,t} \left[ \lambda_i \lambda_j \cdot E(a_i,a_j,a_V,a_A)\right]\,.\label{def-A}
\eeq
Here we have expressed all the quantities by means of the following ratios
\begin{equation}
a_{\alpha} \equiv m^2_{\alpha} /M^2_W \,, (\alpha=i,j) \quad {\rm and}   \quad  
a_v \equiv M^2_v/M^2_W \,, (v=V,A) \ .
\end{equation} 
Indicating with $g_{\rm EW}$ the weak-coupling constant and $\tilde{g}$ the coupling constant governing the massive spin-one self interactions and by expanding up to the order in  ${\cal O}(g^4_{\rm EW}/\tilde{g}^4)$ one can rewrite the previous expression as:
\beq
E(a_i,a_j,a_V,a_A) = E_0(a_i,a_j) + \frac{g^2_{\rm EW}}{\tilde{g}^2} \Delta E(a_i,a_j,a_V,a_A)\,.
\eeq
The explicit expressions can be found in the appendix and are consistent with the results in \cite{Inami:1980fz}. 
The SM contribution is fully contained in $E_0$ and the technicolor one appear first in $\Delta E$. 
The latter can be divided into a vector and an axial-vector contribution as follows:
\beq
\Delta E (a_i,a_j,a_V,a_A) 
=
h(a_i,a_j,a_V) + (1 - \chi)^2 \cdot h(a_i,a_j,a_A)\,,
\eeq
where the expressions for 
$h(a_i,a_j,a_v)$ are reported in the appendix. The quantity $\chi$ was introduced first in \cite{Appelquist:1998xf,Appelquist:1999dq}. Subsequently the associated effective Lagrangian  \cite{Appelquist:1998xf,Appelquist:1999dq} was extended to take into account terms involving the space-time $\epsilon^{\mu\nu\rho\gamma}$ tensor, and topological terms, in \cite{Duan:2000dy} for any technicolor models for which the global symmetry group is either $SU(N_f)\times SU(N_f)$ or $SU(2N_f)$ breaking spontaneously respectively to $SU(N_f)$ or $Sp(2N_f)$, and $N_f$ is the number of techniflavors \footnote{We also note that the approach  used in \cite{Duan:2000dy} to add the $\epsilon$ terms is the most general one given that preserves the {\it true} global symmetries of the underlying gauge theory. 
{}For example the interpretation of the new spin-one states as gauge bosons of a local version of the flavor symmetry allows to relate the coefficients of the $\epsilon$ terms to the one of the gauged Wess-Zumin-Witten topological term \cite{Kaymakcalan:1983qq}. This approach is, however, not justified on theoretical grounds.}.

The axial-vector decay constant is  directly proportional to the quantity $(1-\chi)^2$. The vector and axial decay constant are:
\begin{equation}
f_V^2 = \frac{M^2_V}{\tilde{g}^2} \, \qquad f_A^2 =  \frac{M^2_A}{\tilde{g}^2} (1-\chi)^2 \ .
\end{equation}
{}Note also that for $\chi=2$ and $\chi=0$ the vector and axial-vector meson contributions are identical while for $\chi=1$ only the direct technirho contribution survives. The limit $\chi =0 $ and $M_V = M_A =M$ corresponds to the {\it custodial   technicolor} model introduced in  \cite{Appelquist:1998xf,Appelquist:1999dq,Duan:2000dy}. In this limit the $S$-parameter vanishes identically because is protected by a new symmetry.  We also write: 
\beq
A(a_V,a_A)
=
A_0 + \frac{g^2_{\rm EW}}{\tilde{g}^2} \cdot \Delta A(a_V,a_A)\,.
\label{fin-A}
\eeq
Upon taking into account the unitarity of the CKM matrix and setting $a_u \to 0$ one has
\beq
A_0
=
\eta_1 \cdot \lambda^2_c \cdot  \bar{E}_0(a_c) 
+
\eta_2 \cdot \lambda^2_t \cdot \bar{E}_0(a_t)
+
\eta_3 \cdot 2 \lambda_c \lambda_t \cdot \bar{E}_0(a_c,a_t)
\,, \label{A-SM}
\eeq
and 
\beq
\Delta A(a_V,a_A)
=
\eta_1 \cdot \lambda^2_c \cdot  \Delta \bar{E}(a_c,a_V,a_A) 
+
\eta_2 \cdot \lambda^2_t \cdot \Delta \bar{E}(a_t,a_V,a_A)
+
\eta_3 \cdot 2 \lambda_c \lambda_t \cdot \Delta \bar{E}(a_c,a_t,a_V,a_A)
\,,\nonumber \\\label{A-MWT}
\eeq
where $\eta_{1,2,3}$ are the QCD corrections to $\bar{E}_0$ and $\Delta \bar{E}$. The explicit expressions for  the functions  $\bar{E}$ and $\Delta \bar{E}$  various expressions are provided in the appendix. The expressions simplify for the $ \Delta \bar{E}$ in the relevant limit  $a_v \gg a_t,a_c$: 
\beq
\Delta \bar{E}(a_c,a_t,a_V,a_A) &\simeq& 
\left( 7.28 \times 10^{-5} \right) 
\times \left[ \frac{1}{a_V} + \frac{(1-\chi)^2}{a_A} \right]\,, \\[1ex]
\Delta \bar{E}(a_c,a_V,a_A) &\simeq& 
\left( 3.13 \times 10^{-8} \right) \times \left[ \frac{1}{a_V} + \frac{(1-\chi)^2}{a_A} \right]\,\\[1ex]
\Delta \bar{E}(a_t,a_V,a_A) &\simeq& 
\left( -3.30 \right) \times \left[ \frac{1}{a_V} + \frac{(1-\chi)^2}{a_A} \right]\,.
\eeq
De facto, the formulae above are a reasonable approximation for $M_{V,A} > 400 \GeV$. 
The numerical prefactors  are independent of the specific model of dynamical electroweak symmetry breaking but depend on the SM values for $a_c$ and $a_t$.

\section{Minimal Flavor Constraints} 
We can now compare the generic technicolor effects encoded in Eq.(\ref{lag-eff-dF=2}) and due primarily to the techni-vector  and axial vector contributions with
the CP-violation parameter $\epsilon_K$ in the $K^0$ meson system 
as well as the mass difference of the $K^0-{\bar K}^0$ and $B^0_q-{\bar B}^0_q$ mesons  systems with $q=d,s$.

We recall that the absolute value of the CP-violation parameter in the $K^0-\bar{K}^0$ system 
is given by~\cite{Buchalla:1995vs}:
\beq
\left( |\epsilon_K| \right)_{\rm full}
=
\frac{G^2_F M^2_W}{12 \sqrt{2} \pi^2} \times \left[ \frac{M_K}{\Delta M_K} \right]_{\rm exp.} \hspace*{-3ex} 
\times B_K f^2_K \times [ - {\rm Im}A(a_V,a_A) ] \,.
\label{def-epsilon-K}
\eeq
The meson mass difference in the $Q^0-\bar{Q}^0\, ,\, Q=(K,B_d,B_s)$ system is given by
\beq
\left( \Delta M_Q \right)_{\rm full} 
&\equiv& 
2 \cdot \left| \langle \bar{Q}^0 | - {\cal L}^{\Delta F=2}_{\rm eff} | Q^0 \rangle \right|
=
\frac{G^2_F M^2_W}{6\pi^2} \cdot f^2_Q \cdot M_Q \times B_Q\times \left| A(a_V,a_A) \right|
\label{def-massdiff-Q}\,,
\eeq
where 
$f_Q$ is the decay constant of the $Q$-meson and  
$M_Q$ is its mass. $B_Q$ is identified with the QCD bag parameter 
correcting for possible deviations of the true value of 
the matrix elements $\langle \bar{Q}^0 | - {\cal L}^{\Delta F=2}_{\rm eff} | Q^0 \rangle $ 
from its approximate value computed using the vacuum insertion approximation.
This bag parameter is an intrinsic QCD contribution 
and we assume that the technicolor sector does not contribute to the bag parameter~\footnote{This is a particularly good approximation when the technicolor sector does not have techiquarks charged under ordinary color. The best examples are Minimal Walking Technicolor models.  }. 
There are many estimates available for the bag parameters,  such as the ones from the lattice~\cite{Aoki:2005ga,Okamoto:2005zg}, 
$1/N$-approximation~\cite{Bardeen:1987vg}, etc.
In this paper we use, for definitiveness, the values quoted in \cite{Nierste:2009wg}.
The experimental values of $G_F,M_W,f_Q,M_Q,\Delta M_Q$ 
and the bag parameter $B_Q$ are shown in Table~\ref{mass-difference-VI}.

\begin{table}
\begin{center}
{
\begin{tabular}{|c||c c||c|}
\hline
$G_F$  & $1.1664 \times 10^{-5}$ & $\GeV^{-2}$ &Ref.~\cite{Amsler:2008zzb}\\ \hline
$M_W$ & $80.398$ & $\GeV$ & Ref.~\cite{Amsler:2008zzb} \\ \hline 
$m_t$ & $161.3 \pm 1.8$ & $\GeV$ & $\overline{\rm MS}$ mass in Ref.~\cite{Amsler:2008zzb} \\ \hline
$m_c$ & $1.274 ^{+0.036}_{-0.045}$ & $\GeV$ & $\overline{\rm MS}$ mass in Ref.~\cite{Amsler:2008zzb} 
\\ \hline\hline
$M_K$  & $497.61 \pm 0.02$ & $\MeV$ & Ref.~\cite{Amsler:2008zzb} \\ \hline
$\Delta M_K$ & 
$5.292 \pm 0.0009$ & ${\rm ns}^{-1}$
&  Ref.~\cite{Amsler:2008zzb} \\ \hline
$|\epsilon_K|$ & $(2.229 \pm 0.012) \times 10^{-3}$ & \ & Ref.~\cite{Amsler:2008zzb} \\ \hline 
$f_K$   & $155.5$ & $\MeV$ & Ref.~\cite{Amsler:2008zzb} \\ \hline 
$B_K$ & $0.72 \pm 0.040 $ & \ & Ref.~\cite{Nierste:2009wg} \\ \hline\hline
$M_{B_d}$ & $5279.5 \pm 0.3$ & $\MeV$ & Ref.~\cite{Amsler:2008zzb} \\ \hline 
$\Delta M_{B_d}$ & 
$0.507 \pm 0.005$ & ${\rm ps}^{-1}$ 
& Ref.~\cite{Amsler:2008zzb} \\ \hline 
$f_{B_d}\sqrt{B_{B_d}} $ & $225 \pm 35$ & $\MeV$ & Ref.~\cite{Nierste:2009wg} \\ \hline \hline
$M_{B_s}$ & $5366.3 \pm 0.6$ & $\MeV$ & Ref.~\cite{Amsler:2008zzb} \\ \hline 
$\Delta M_{B_s}$ & 
$17.77 \pm 0.10$ & ${\rm ps}^{-1}$
& Ref.~\cite{Amsler:2008zzb} \\ \hline 
$f_{B_s} \sqrt{B_{B_s}}$ & $270 \pm 45$ & $\MeV$ & Ref.~\cite{Nierste:2009wg} \\ \hline
\end{tabular}
}
\end{center}
\caption{Fermi constant $(G_F)$, $W^\pm$ boson mass $(M_W)$, 
                  top quark and charm quark masses in the $\overline{\rm MS}$-scheme $(m_t,m_c)$, 
                  meson masses $(M_K,M_{B_q})$, indirect CP violation parameter $(\epsilon_K)$,
                  meson mass difference $(\Delta M_K,\Delta M_{B_q})$, 
                  decay constants $(f_K,f_{B_q})$ and Bag parameters $(B_K,B_{B_q})$.
                  We show the central value for $G_F,M_W,f_K$.
                  \label{mass-difference-VI} }
\end{table}
It is convenient to define the following quantities:
\beq
\delta_\epsilon 
&\equiv 
&
\frac{g^2_{\rm EW}}{\tilde{g}^2} \cdot 
\frac{{\rm Im}\Delta A(a_V,a_A)}{{\rm Im}A_0}\,,
\label{def-CPviolation-K}\\[2ex]
\delta_{M_Q} 
&\equiv 
&
\frac{g^2_{\rm EW}}{\tilde{g}^2} \cdot 
\frac{\Delta A(a_V,a_A)}{A_0}\,.
\label{def-mass-difference}
\eeq
Using these expressions we write $(|\epsilon_K|)_{\rm full}$ and $(\Delta M_Q)_{\rm full}$ as
\beq
\left( |\epsilon_K| \right)_{\rm full}= \left( |\epsilon_K| \right)_{\rm SM} \times (1 + \delta_\epsilon) 
\quad ,\quad
\left( \Delta M_Q \right)_{\rm full}= \left( \Delta M_Q \right)_{\rm SM} \times \left| 1 + \delta_{M_Q} \right|\,.
\label{CP-and-dmass-full}
\eeq
Of course, $\left( |\epsilon_K| \right)_{\rm SM}$ and $\left( \Delta M_Q \right)_{\rm SM}$ 
are the SM expressions encoded in: 
\beq
\left( |\epsilon_K| \right)_{\rm SM}
&=&
\frac{G^2_F M^2_W}{12 \sqrt{2} \pi^2} \times \left[ \frac{M_K}{\Delta M_K} \right]_{\rm exp.} \hspace*{-3ex}
\times B_K f^2_K \times [ - {\rm Im}A_0 ]\,,\\[1ex]
\left( \Delta M_Q \right)_{\rm SM}  
&=&
\frac{G^2_F M^2_W}{6\pi^2} \cdot f^2_Q \cdot M_Q \times B_Q\times | A_0 |\,.
\label{CP-and-dmass-SM}
\eeq
They assume the values:  
\beq
\left( |\epsilon_K| \right)_{\rm SM} &=& (2.08^{+0.14}_{-0.13}) \times 10^{-3}\,, 
\label{epsilonK-SM}\\[1ex]
\left( \Delta M_K \right)_{\rm SM}&=& (3.55^{+1.09}_{-1.00})  \hspace*{3ex} {\rm ns}^{-1} \,,
\label{deltaMK-SM}\\[1ex]
\left( \Delta M_{B_d} \right)_{\rm SM}&=& (0.56^{+0.19}_{-0.16}) \hspace*{3ex} {\rm ps}^{-1}\,,\,
\label{deltaMBd-SM}\\[1ex]
\left( \Delta M_{B_s} \right)_{\rm SM}&=& (17.67^{+6.38}_{-5.40}) \hspace*{2ex} {\rm ps}^{-1}
\label{deltaMBs-SM}\,.
\eeq
To evaluate the expressions above we used  the values of $G_F,M_W,M_Q,\Delta M_K,f_Q,B_Q$ in Table~\ref{mass-difference-VI}. We also used the 
 CKM matrix elements expressed in the Wolfenstein parameterization \cite{Wolfenstein:1983yz} and reported in Appendix B. 
We also need the QCD correcting factors $\eta_{1,2,3}$ to evaluate $A_0$. Following \cite{Nierste:2009wg} these are:
\beq
\eta_1 = (1.44 \pm 0.35)\cdot \left( \frac{1.3 \GeV}{m_c} \right)^{1.1} \quad , \quad 
\eta_2 = 0.57 \quad , \quad
\eta_3 = 0.47 \pm 0.05\,,
\eeq
for the kaon system while we also need $\eta_B = 0.55$ \cite{Nierste:2009wg}, corresponding to $\eta_2$, for the system containing a bottom quark.

Given that  
\beq
\bar{E}_0(a_c) \simeq 2.51 \times 10^{-4} \quad,\quad
\bar{E}_0(a_t) \simeq 2.27 \quad,\quad
\bar{E}_0(a_c,a_t) \simeq 2.22 \times 10^{-3}\,,
\label{IL-values}
\eeq
and that the CKM derived quantities $\lambda^2_c,\lambda^2_t,\lambda_c\lambda_t$ are roughly of the same order for the $B^0_q$ system we neglected the $\eta_1  \bar{E}_0(a_c)$ and $\eta_3   \bar{E}_0(a_c,a_t)$ terms when providing the estimates for  this system. 
The uncertainty in  Eqs~(\ref{epsilonK-SM})  -(\ref{deltaMBs-SM}) were deduced 
by propagating the theoretical ones plaguing  $\eta_{1,3}$, $B_K,f_{B_q}$ and $\sqrt{B_{B_q}}$~\cite{Buchalla:1995vs,Nierste:2009wg}.

We are now ready to compare the SM value given in Eq.(\ref{epsilonK-SM}) with the experimental one in Table~\ref{mass-difference-VI} and read off the
constrain on $\delta_\epsilon$ which is:
{\color{black}
\beq
\delta_\epsilon  =  \left( 7.05^{+ 7.93}_{-7.07} \right) \times 10^{-2} \quad (68\%\cl) \,.\label{constraint-MWT-eK}
\eeq
}
In order to compare the corrections associate to the kaon mass $\Delta M_K$ we formally separate the short distance contribution from the long distance one and write
\beq
\Delta M_K = \left( \Delta M_K \right)_{\rm SD} + \left( \Delta M_K \right)_{\rm LD}\,.
\eeq
Here $\left( \Delta M_K \right)_{\rm SD}$ encodes the short distance contribution which must be confronted with the technicolor one
 Eq.(\ref{def-mass-difference}). The SM contribution to the short distance kaon mass difference evaluated in Eq.(\ref{deltaMK-SM}) is circa $70 \%$ of $(\Delta M_K)_{\rm exp}$.
The long distance contribution, $\left( \Delta M_K \right)_{\rm LD}$, 
corresponds to the exchange of the light pseudoscalar mesons 
and its contribution may yield the remaining $30 \%$ of the experimental value $\left( \Delta M_K\right)_{\rm exp.}$
~\cite{Buchalla:1995vs}.
However,  it is difficult to pin-point the $\left( \Delta M_K \right)_{\rm LD}$ contribution 
~\cite{Buchalla:1995vs,Donoghue:1992dd} and hence we can only derive very weak constraints  from $\delta_{M_K}$. In fact we simply require that 
\begin{equation}
\left( \Delta M_K \right)_{\rm SD} = \left( \Delta M_K \right)_{\rm SM} \,  |1 + \delta_{M_K}| 
\leq \left( \Delta M_K \right)_{\rm exp.} \ .
\end{equation}
This means that:
\beq
|1 + \delta_{M_K}|  \leq 2.08 \quad (68\%\cl) \,. \label{constraint-MWT-dMK}
\eeq
On the other hand  the short distance contribution dominates the $B^0_q-\bar{B}^0_q$ mass difference~\cite{Donoghue:1992dd} yielding the following constraints:
{\color{black}
\beq
 |1 + \delta_{M_{Bd}}|  &=& 0.91^{+0.38}_{-0.24} \quad (68\%\cl) \,,\label{constraint-MWT-dMBd}\\[2ex]
 |1 + \delta_{M_{Bs}}|  &=& 1.01^{+0.44}_{-0.27} \quad (68\%\cl) \,.  \label{constraint-MWT-dMBs}
\eeq
}

These constitute the {\it minimal}  flavor constraints on any model of dynamical electroweak symmetry breaking. On the top of these corrections one has the ones coming from a given explicit extended technicolor model. Typically these models are hard to construct and, hence, to constrain. On the other hand assuming the existence of a successful extension, meaning that it provides the correct masses to the SM fermions and no direct flavor changing neutral currents effects, one has still to consider the experiment constraints above on the technicolor sector we have just computed.

{\color{black}
Although the analytic formulae for $\Delta A$ are valid for any value assumed by the vector meson masses they simplify considerably in the limit $M^2_V,M^2_A \gg m^2_t,m^2_c$. We term it the intermediate vector limit, iVL, and define the associated quantities with $\delta^{({\rm iVL})}_\epsilon,\delta^{({\rm iVL})}_{M_Q}$. They read:
\beq
\delta^{({\rm iVL})}_\epsilon \simeq  -2.14 \times W\quad , \quad 
\delta^{({\rm iVL})}_{M_K} \simeq -0.026 \times W\quad , \quad
\delta^{({\rm iVL})}_{B_d} = \delta^{({\rm iVL})}_{B_s} \simeq -2.90 \times W\,,
\label{MFconstraint-app}
\eeq 
where the numerical values depend on $a_c\,,a_t\,,\eta_{1,2,3}\,,\eta_B\,,\lambda_i$ with
\begin{equation} 
W=\frac{g^2_{\rm EW}}{2\tilde{g}^2} \left[ \frac{1}{a_V} + \frac{(1-\chi)^2}{a_A} \right] \ , \quad W \equiv  \frac{g^2_{\rm EW} M^2_W}{2} \left[\Pi''_{33}(0)\right] \ , 
\end{equation} 
and $\Pi_{33}$ the $W_3W_3$ corrections to the vacuum polarization due to the exchange of the new heavy vectors.   
The $Y$ parameter is defined as   \cite{Barbieri:2004qk} 
\begin{equation}
Y \equiv \frac{g'^2_{\rm EW} M^2_W}{2} \left[\Pi''_{BB}(0)\right]  
\end{equation} 
and for a generic minimal model of technicolor, i.e. in which the techniquarks are not charged under ordinary color interactions, we have: 
\beq
Y=\frac{g'^2_{\rm EW}}{2\tilde{g}^2} \left[ \frac{1+4y^2}{a_V} + \frac{(1-\chi)^2}{a_A} \right]\,.
\eeq
The flavor constraints on the $W$ parameter, as we shall see, are important and will provide tight constraints on the underlying technicolor dynamics, or alike models. We will then compare the limits with the ones deriving from LEP II data, i.e. 
$W=(-0.2\pm0.8)\times 10^{-3}\,$ and $Y=(0.0\pm1.2)\times 10^{-3}$
corresponding to the $68\%\cl$ constraints for a heavy Higgs in \cite{Barbieri:2004qk}. As a consistency check  one can see that the expression for $W$ and $Y$ coincide with the ones derived in \cite{Foadi:2007se}. 
}

\section{Constraining Models of Dynamical Electroweak Symmetry Breaking}
We will now use the minimal flavor experimental information to reduce the parameter space of a general class of models of dynamical electroweak symmetry breaking.

\subsection{Scaled up version of QCD:  The running case }
If the underlying TC theory is QCD like we can impose the standard 1st and 2nd Weinberg's sum rules as shown in  \cite{Appelquist:1998xf,Appelquist:1999dq,Duan:2000dy,Foadi:2007ue}
\beq
\text{1st WSR} &:& f^2_V- f^2_A = f^2_\pi = \left( \frac{v_{\rm EW}}{\sqrt{2}}\right)^2 \,,
\label{1stWSR}\\[2ex]
\text{2nd WSR} &:& f^2_V \, M^2_V - f^2_A \, M^2_A = 0\,.
\label{2ndWSR-QCDTC}
\eeq
with $f_V$ and $f_A$ the vector and axial decay constants. One obtains exactly the expression in \cite{Foadi:2007ue} via the re-definition $F_i = \sqrt{2} f_i\,,(i=V,A,\pi)$. Using the explicit expressions of the decay constants in terms of the coupling $\tilde{g}$ and vector masses provided in \cite{Appelquist:1998xf,Appelquist:1999dq,Duan:2000dy,Foadi:2007ue} and imposing the above sum rules we derive:
\beq
\frac{1}{a_V}
= \frac{g^2_{\rm EW} S}{16\pi} -  \frac{1}{a_A}  \quad ( \geq 0 )\, ,
\label{aV-aAfromQCDWSR}
\eeq
with the $S$-parameter~\cite{Peskin:1991sw} reading \cite{Appelquist:1998xf,Appelquist:1999dq,Duan:2000dy,Foadi:2007ue}: 
\beq
S 
\equiv 8\pi \left[ \frac{f^2_V}{M^2_V} - \frac{f^2_A}{M^2_A} \right]
= \frac{8 \pi}{\tilde{g}^2} \,\left[ 1 - (1 - \chi)^2 \right]\,.
\label{def-S}
\eeq
The condition above yields the following additional constraint for $\tilde{g}$ by simply noting that the quantity $(1-\chi)^2$  is positive: 
\beq
\tilde{g} < \sqrt{\frac{8\pi}{S}} \,.
\label{GHLS-g-constraint}
\eeq
The constraints on $(M_A, \tilde{g})$ induced by (\ref{aV-aAfromQCDWSR}) and (\ref{GHLS-g-constraint}) are stronger, for a given $S$, than the ones deriving from flavor experiments
and expressed in (\ref{constraint-MWT-eK})-(\ref{constraint-MWT-dMBs}). This is not surprising given that in an ordinary technicolor theory the spin-one states are very heavy. However the situation changes when allowing for a walking behavior. 

\subsection{Walking Models}
Besides the flavor constraints one has also the ones due to the electroweak precision measurements  \cite{Foadi:2007se} as well as the unitarity constraint of $W_L-W_L$ scattering \cite{Foadi:2008xj}. We will consider all of them. As for the technicolor case we reduce the number of independent parameters at the effective Lagrangian level via the 1st- and the second modified \cite {Appelquist:1998xf}  Weinberg sum rules which now read 
~\cite{Appelquist:1998xf}
\beq
\text{1st WSR} &:& f^2_V- f^2_A = f^2_\pi = \left( \frac{v_{\rm EW}}{\sqrt{2}}\right)^2 \,,
\label{1stWSR-MWT}\\[2ex]
\text{2nd MWSR} &:& f^2_V \, M^2_V - f^2_A \, M^2_A = a \cdot \frac{16 \pi^2}{d({\rm R})} f^4_\pi\,,
\label{2ndWSR-MWT}
\eeq 
where 
$a$ is a number expected to be positive and ${\cal O}(1)$~\cite{Appelquist:1998xf}. 
$d({\rm R})$ is the dimension of the representation of the underlying technifermions as shown in \cite{Foadi:2007ue}.  We have now: 
\beq
M^2_A 
< \frac{8\pi f^2_\pi}{S} \frac{2-\chi}{1-\chi} 
= \frac{8\pi f^2_\pi}{S} \left[ 1 + \frac{1}{\sqrt{1-\frac{\tilde{g}^2 S}{8\pi}}}\right]\,.
\label{MA-constraint}
\eeq

In Figure~\ref{plot-constraints-walking},
we show the allowed region in the $(M_A,\tilde{g})$-plane after having imposed the minimal flavor constraints  due to the experimental values of  $|\epsilon_K|$ and $\Delta M_Q$  obtained using 
Eqs.~(\ref{constraint-MWT-eK})-(\ref{constraint-MWT-dMBs})
together with the theoretical constraints for $\tilde{g},M^2_A$ obtained via Eqs.~(\ref{GHLS-g-constraint})  and (\ref{MA-constraint}).
To obtain Fig.~\ref{plot-constraints-walking}, 
we used the expressions for $A_0$ and $\Delta A(a_V,a_A)$ shown respectively in \eqref{A-SM} and \eqref{A-MWT} in which
 $a_V = M^2_V/M^2_W$ reads:
\beq
a_V = \left[ 1-\frac{\tilde{g}^2}{8\pi} S \right] \cdot a_A + \frac{2 \tilde{g}^2}{g^2_{\rm EW}} \,.
\label{aV-aA-from1stWSR}
\eeq
We obtained the last expression imposing the first Weinberg sum rule of Eq.~(\ref{1stWSR-MWT}).  
Given that the upper bound for $\delta_{M_K}$ is always larger than the theoretical estimate in the region $M_A>200\GeV$ we conclude that 
the  $\Delta M_K$ constraint is not yet very severe and hence it is not displayed in Fig.~\ref{plot-constraints-walking}.

To make the plots we need also the value of the $S$ parameters and hence we analyzed as explicit example minimal walking technicolor models. 

\subsubsection{Minimal Walking Technicolor}
For definitiveness we use for $S$ the naive MWT estimate,  i.e. $S=1/(2\pi)$~\cite{Sannino:2004qp,Dietrich:2006cm} while 
$\tilde{g}$ is constrained via Eq.~(\ref{GHLS-g-constraint}) to be $\tilde{g} < 12.5$.

We have plotted the various constraints on the $(M_A,\tilde{g})$-plane for MWT in the upper and lower left panel of Fig.~\ref{plot-constraints-walking}.  In the upper  (lower) left figure we compare the $68\%\cl$  ($95\%\cl$) allowed regions coming from the minimal flavor constraints (the darker region above the blue-dotted line) with the ones from LEP II data (region above the green-dashed line).  It is clear that the flavor constraints are stronger for the $68\%\cl$ case but are weaker for the $95\%\cl$ one with respect to the constraints from LEP II data. 

The region above the straight solid line is forbidden by the condition $\tilde{g} < 12.5$ while the region below the black solid  curve (on the right corner) by the condition (\ref{MA-constraint}) \footnote{It might be that the original theory behind Minimal Walking Technicolor, i.e. the $SU(2)$ gauge theory with two Dirac flavors transforming according to the adjoint representation of the gauge group, develops an infrared fixed point. This case seems to be supported by recent lattice results \cite{Catterall:2007yx,Catterall:2008qk,DelDebbio:2008zf,Hietanen:2008mr,Hietanen:2009az}. Such a possibility was first predicted in \cite{Sannino:2008ha} using the  the beta function proposed in \cite{Ryttov:2007cx}. The Ryttov-Sannino (RS) beta function results have, very recently, been shown to be consistent with new solutions of the 't Hooft anomaly matching conditions presented in \cite{Sannino:2009qc}. Other analytic analysis further supporting the RS beta function have recently appeared \cite{Poppitz:2009uq,Armoni:2009jn}. If the presence of such a nontrivial fixed point is confirmed this will not reduce the phenomenological interest for this theory but rather increase it both for constructing interesting new models of dynamical electroweak symmetry breaking \cite{AST} or  to provide {\it minimal models of unparticle physics} similar to the ones suggested in \cite{Sannino:2008nv}. }. 

\subsubsection{Next to Minimal Walking Technicolor (NMWT)}
\begin{figure}[h!]
\begin{center}
\includegraphics[width=0.4\textwidth,height=0.4\textwidth]{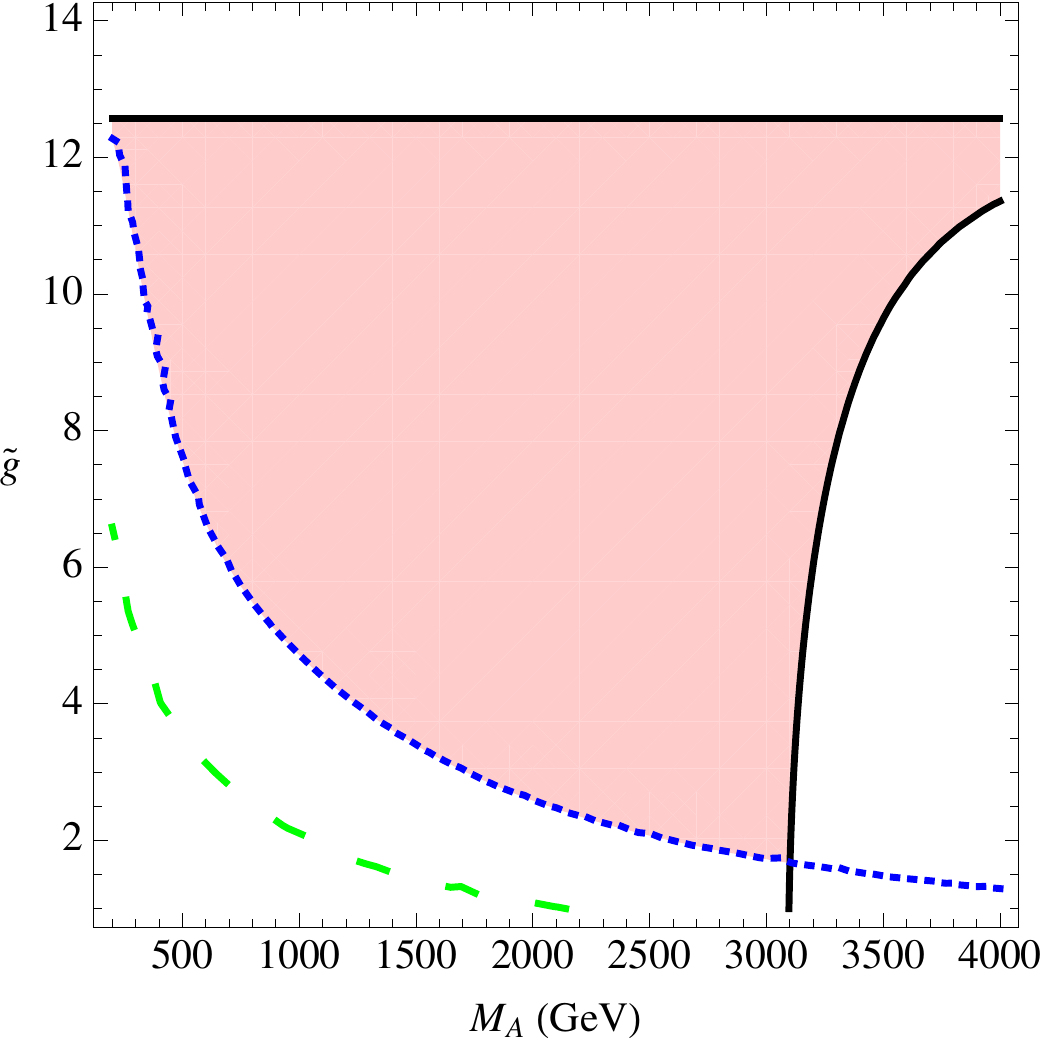} \hspace*{5ex}
\includegraphics[width=0.4\textwidth,height=0.4\textwidth]{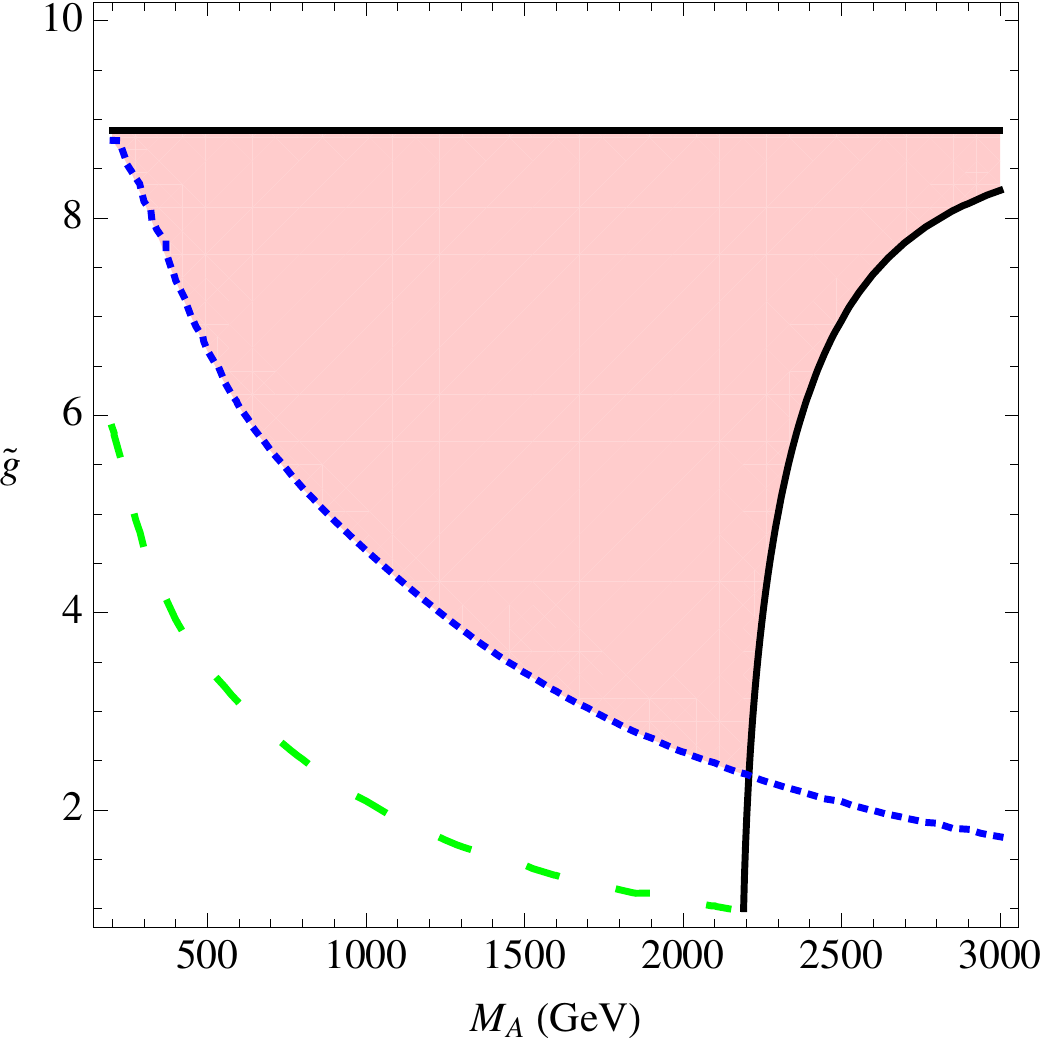}\\[3ex]
\includegraphics[width=0.4\textwidth,height=0.4\textwidth]{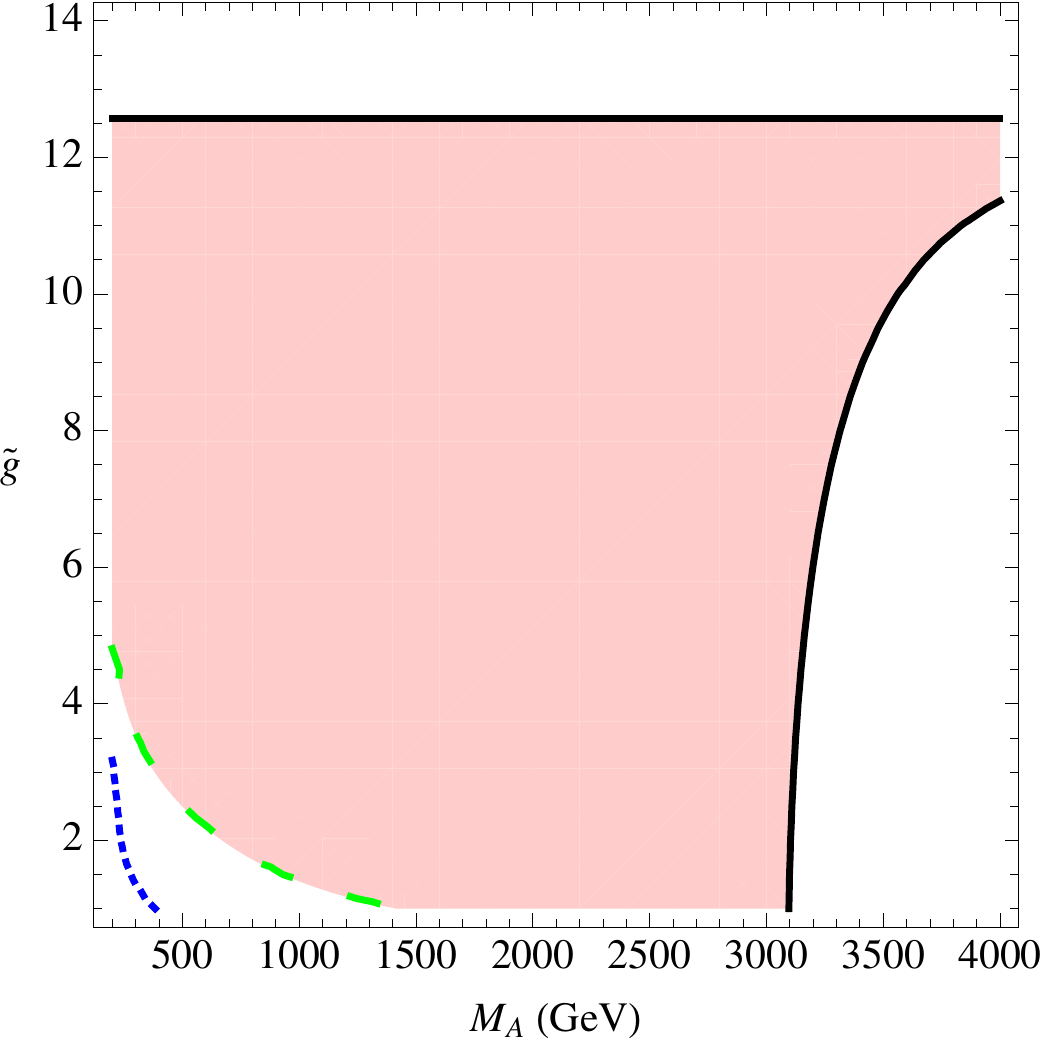} \hspace*{5ex}
\includegraphics[width=0.4\textwidth,height=0.4\textwidth]{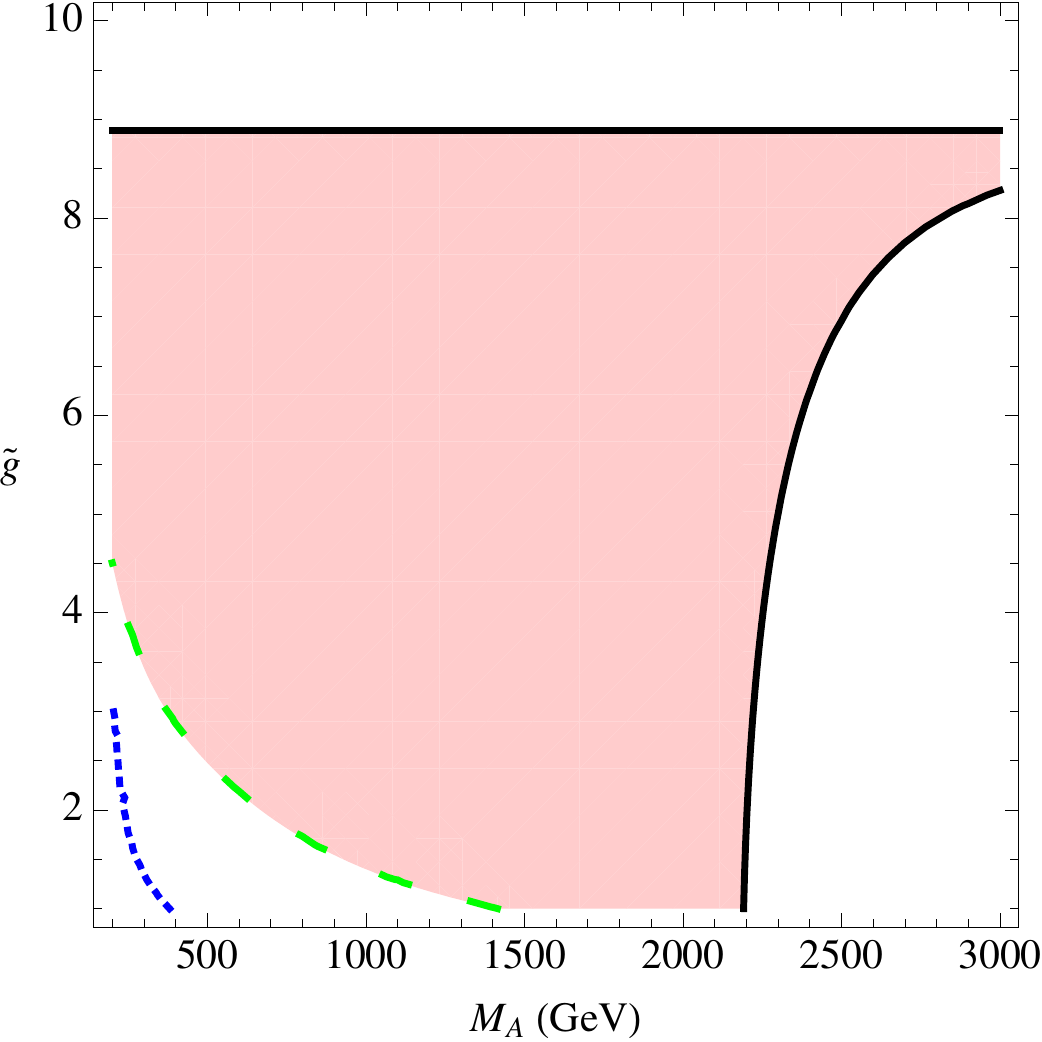}
\caption{The upper and lower left panels represents the allowed region in the $(M_A,\tilde{g})$-plane for MWT respectively for the $68\%\cl$ and  $95\%\cl$. A similar analysis is shown for NMWT in the right hand upper and lower panels. 
The region above the straight solid line is forbidden by the condition $\tilde{g} < 12.5$  for MWT and  $\tilde{g} < 8.89$ for NMWT while the region below the black solid  curve (on the right corner) by the condition (\ref{MA-constraint}).
In the two upper (lower) plots the blue dotted lines correspond to the $68\%\cl$ ($95\%\cl$) flavor constraints while the green dashed lines are the $68\%\cl$ ($95\%\cl$)   from LEP II data. 
The flavor constraints come only from $\epsilon_K$ since the ones from $\Delta M_{B_q}$ are not as strong.  }                  
\label{plot-constraints-walking}
\end{center}
\end{figure}

In this case the naive $S$ is approximately $1/\pi$~\cite{Sannino:2004qp,Dietrich:2006cm} and the constraint on 
$\tilde{g}$ from  Eq.~(\ref{GHLS-g-constraint}) yields  $\tilde{g} < 8.89$. We have plotted the various constraints on the $(M_A,\tilde{g})$-plane for NMWT  in the upper and lower right panel of Fig.~\ref{plot-constraints-walking} for the $68\%\cl$ and $95\%\cl$  constraints. We see again that the flavor constraints are stronger for the $68\%\cl$ case but are weaker for the $95\%\cl$ one. 

\subsection{Custodial Technicolor}
\begin{figure}[h!]
\begin{center}
\includegraphics[width=0.4\textwidth,height=0.4\textwidth]{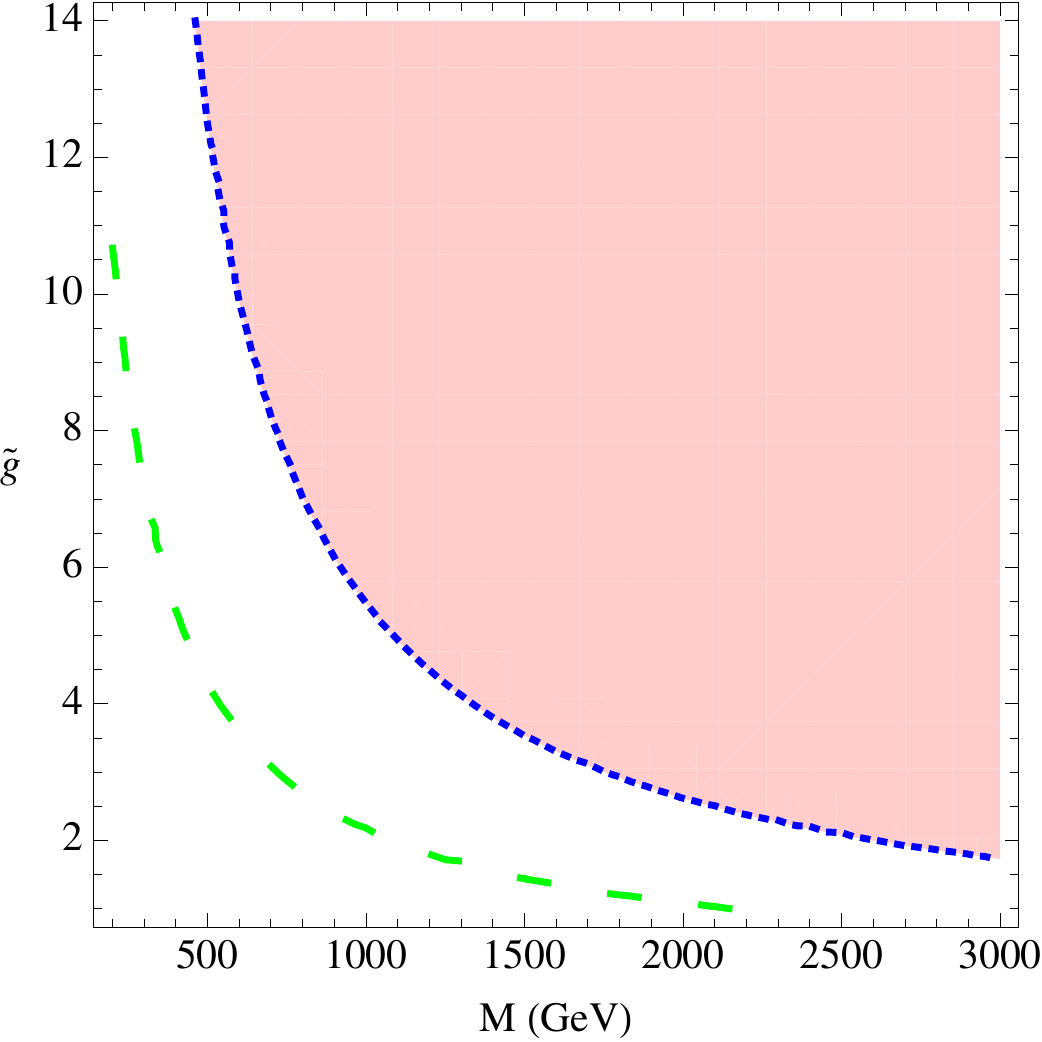} \hspace*{5ex}
\includegraphics[width=0.4\textwidth,height=0.4\textwidth]{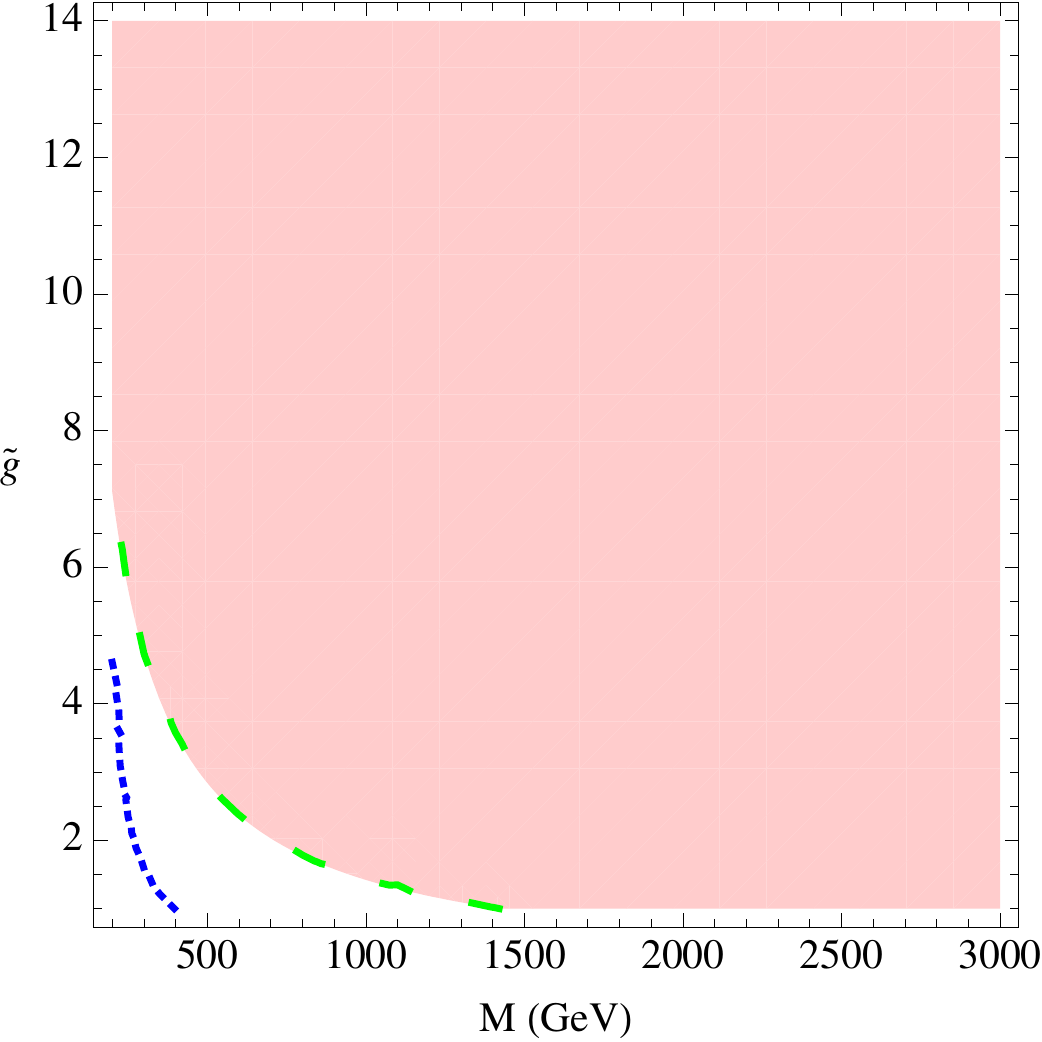}
\caption{The left (right) panel represents the allowed region in the $(M_A,\tilde{g})$-plane for CT respectively for the $68\%\cl$ ($95\%\cl$). 
In the two upper (lower) plots the blue dotted lines correspond to the $68\%\cl$ ($95\%\cl$) flavor constraints while the green dashed lines are the $68\%\cl$ ($95\%\cl$)   from LEP II data. 
The flavor constraints come only from $\epsilon_K$ since the ones from $\Delta M_{B_q}$ are not as strong.  }                  
\label{plot-contraints-CT}
\end{center}
\end{figure}

In the limit $M_A= M_V = M$ and $\chi =0$ the effective theory acquires 
 a new symmetry  \cite{Appelquist:1999dq,Duan:2000dy}. This new symmetry relates a vector and an axial field and can be shown to work as a custodial symmetry for the $S$ parameter \cite{Appelquist:1999dq,Duan:2000dy}.  
 The only non-zero electroweak parameters are now: 
 \begin{equation}
 W = \frac{g^2_{\rm EW}}{\tilde{g}^2} \frac{M^2_W}{M^2} \ , \qquad Y = \frac{{g^{\prime}_{EW}}^2}{2\tilde{g}^2} \frac{M^2_W}{M^2} (2 + 4 y^2 ) \ .
 \end{equation}
It was already noted in \cite{Foadi:2007se} that a custodial technicolor model cannot be easily achieved via an underlying walking dynamics and should be interpreted as an independent framework. This is so since custodial technicolor models do not respect the Weinberg's sum rules \footnote{One can, of course, imagine more complicated vector spectrum leading to such a symmetry.}.  This symmetry is also present in the BESS models \cite{Casalbuoni:1988xm,Casalbuoni:1995yb,Casalbuoni:2007dk} which will, therefore, be constrained as well. We directly compare in the Fig.~\ref{plot-contraints-CT} the constraints on the custodial technicolor parameter region ($M$,$\tilde{g}$) coming from LEP II and flavor constraints and find a similar trend as for the other cases. 

 \section{Summary}

Flavor constraints are  relevant for models of dynamical electroweak symmetry breaking  with light spin-one resonances, in fact,  any model featuring spin-one resonances with the same quantum numbers of the SM gauge bosons will have to be confronted with these flavor constraints.  Combining the flavor and LEP II data the new value for $W$ at the one sigma level is  $W_{\rm avg} \simeq  (-1.6^{+ 3.7}_{-3.3}) \times 10^{-2}$. 
\acknowledgments
We thank Mads T. Frandsen, Gino Isidori, Matti J\"arvinen,  Maurizio Pierini, Luca Silvestrini, Koichi Yamawaki and Roman Zwicky for valuable comments and relevant discussions.

\appendix

\section{Relevant Expressions}
We provided the explicit form of each quantity introduced in the main text starting with:
\beq
E_0(a_i,a_j)
&=&
-\frac{3}{4} h_0(a_i,a_j) 
- \frac{7}{4}\left( \frac{1}{a_i-1} + \frac{1}{a_i-1} + 1\right) \nonumber\\[1ex]
&&
+
\left[ \frac{a_i a_j}{a_i-a_j} \left( \frac{1}{4} - \frac{3}{2(a_i-1)}\right) + \frac{7 a_i}{4(a_i-1)^2}\right] \ln a_i
\nonumber \\[1ex]
&&
+
\left[ \frac{a_i a_j}{a_j-a_i} \left( \frac{1}{4} - \frac{3}{2(a_j-1)}\right) + \frac{7 a_j}{4(a_j-1)^2}\right] \ln a_j\,,
\eeq
and
\beq
h_0(a_i,a_j)
=
\frac{1}{a_i - a_j}  
\left[ 
\left( \frac{a_i}{a_i-1}\right)^2 \ln a_i - \left( \frac{a_j}{a_j-1}\right)^2 \ln a_j 
- \frac{1}{a_i-1} + \frac{1}{a_j-1}
\right]\,.
\eeq
We also have: 
\beq
\Delta E (a_i,a_j,a_V,a_A) 
=
h(a_i,a_j,a_V) + (1 - \chi)^2 \cdot h(a_i,a_j,a_A)\,,
\eeq
where 
$h(a_i,a_j,a_v)$ is given by 
\beq
h(a_i,a_j,a_v)
=
&&
\frac{a^2_i \ln a_i}{(a_i-1)^3(a_i-a_j)(a_i-a_v)} 
\cdot 
\left[ a^2_i -\frac{3}{4}  a^2_i a_j - a_i a_j \right]
\nonumber\\
&&
+
\frac{a^2_j \ln a_j}{(a_j-1)^3(a_j-a_i)(a_j-a_v)}
\cdot 
\left[ a^2_j -\frac{3}{4} a_i a^2_j - a_i a_j\right]
\nonumber\\
&&
+
\frac{a^2_v \ln a_v}{(a_v-1)^3(a_v-a_i)(a_v-a_j)} 
\cdot 
\left[ a^2_v -\frac{3}{4} a_i a_j a_v - a_i a_j\right]
\nonumber\\[1ex]
&&
- 
\frac{1}{(a_i-1)(a_j-1)(a_v-1)} 
\cdot 
\left[ \frac{1}{a_i-1} + \frac{1}{a_j-1} + \frac{1}{a_v-1} \right]
\cdot
\left[ 1-\frac{7}{4} a_i a_j\right]
\nonumber\\[1ex]
&&
- 
\frac{1}{(a_i-1)(a_j-1)(a_v-1)} 
\cdot 
\left[ \frac{7}{2} - \frac{27}{8} a_i a_j \right] \,.
\label{VD-contribution-IL}
\eeq
We can now provide the full expression for $A(a_V,a_A)$ 
\beq
A(a_V,a_A)
=
A_0 + \frac{g^2_{\rm EW}}{\tilde{g}^2} \cdot \Delta A(a_V,a_A)\,. 
\label{fin-AA}
\eeq
Taking into account the unitarity constraint from the CKM matrix and setting $a_u \to 0$ one finds: 
\beq
A_0
=
\eta_1 \cdot \lambda^2_c \cdot  \bar{E}_0(a_c) 
+
\eta_2 \cdot \lambda^2_t \cdot \bar{E}_0(a_t)
+
\eta_3 \cdot 2 \lambda_c \lambda_t \cdot \bar{E}_0(a_c,a_t)
\,, \label{AA-SM}
\eeq
and 
\beq
\Delta A(a_V,a_A)
=
\eta_1 \cdot \lambda^2_c \cdot  \Delta \bar{E}(a_c,a_V,a_A) 
+
\eta_2 \cdot \lambda^2_t \cdot \Delta \bar{E}(a_t,a_V,a_A)
+
\eta_3 \cdot 2 \lambda_c \lambda_t \cdot \Delta \bar{E}(a_c,a_t,a_V,a_A) \ .\nonumber \\
\label{AA-MWT}
\eeq
where 
$\Delta \bar{E}(a_i,a_j,a_V,a_A) = \Delta \bar{E}(a_i,a_j,a_v) + (1-\chi)^2 \Delta \bar{E}(a_i,a_j,a_A)$, etc.
$\eta_{1,2,3}$  encodes the QCD corrections for $\bar{E}_0$ and $\Delta \bar{E}$.

Here, $\bar{E}_0(a_i,a_j)$ is given by 
\beq
\bar{E}_0(a_i,a_j)
&=&
\lim_{a_u \to 0} 
\left[ E_0(a_i,a_j) - E_0(a_u,a_j) - E_0(a_i,a_u) + E_0(a_u,a_u) \right]
\nonumber\\[1ex]
&=&
\frac{a_i a_j}{a_i-a_j} 
\left[ K_0(a_i) - K_0(a_j)\right]\,,
\label{Ebar2-SM}
\eeq
with
\beq
K_0(x)
=
\left[ \frac{1}{4} - \frac{3}{2(x-1)} - \frac{3}{4(x-1)^2} \right] \ln x
+
\frac{3}{4(x-1)} \,.
\eeq
We also have:
{\color{black}  
\beq
\Delta \bar{E}(a_i,a_j,a_v) 
&=&
\lim_{a_u \to 0} 
\left[ \Delta E(a_i,a_j,a_v) - \Delta E(a_u,a_j,a_v) - \Delta E(a_i,a_u,a_v) + \Delta E(a_u,a_u,a_v) \right]
\nonumber\\[1ex]
&=&
\frac{a_i a_j \cdot \left[ (a_j -a_v) K(a_i)+ (a_v -a _i) K(a_j) + (a_i - a_j) K(a_v)\right]}{(a_i - a_j)(a_j - a_v)(a_v - a_i)} 
\,,
\label{Ebar2-VD}
\eeq
}
where
\beq
K(x) 
&=&
\frac{- \ln x}{(x-1)^3} 
\cdot \left[ x^2  - \frac{3}{4} a_i a_j \cdot x - a_i a_j \right] \nonumber\\[1ex]
&&+
\frac{1}{x-1} 
\left[ 1-\frac{7}{4} a_i a_j\right] 
\left[ \frac{1}{a_i-1} + \frac{1}{a_j-1} + \frac{1}{a_v-1}\right]
+ \frac{1}{x-1} \left[ \frac{3}{2} + \frac{1}{8} a_i a_j \right]\,.
\label{def-K}
\eeq
Moreover
\beq
\bar{E}_0(a_i)
&\equiv&
\lim_{a_j \to a_i} \bar{E}_0(a_i,a_j) \nonumber\\[1ex]
&=&
\frac{3}{2} \left( \frac{a_i}{a_i-1}\right)^3 \ln a_i
+
\left[ \frac{1}{4} - \frac{9}{4(a_i-1)} -\frac{3}{2(a_i -1)^2}\right] a_i \,,
\label{Ebar1-SM}
\eeq
\beq
\Delta \bar{E}(a_i,a_V,a_A) 
\equiv 
\lim_{a_j \to a_i} \Delta \bar{E}(a_i,a_j,a_V,a_A)
=
\Delta \bar{E}(a_i,a_V) + (1-\chi)^2 \Delta \bar{E} (a_i,a_A)
\,,
\eeq
and 
\beq
\Delta \bar{E}(a_i,a_v) 
&\equiv& \lim_{a_j \to a_i} \Delta \bar{E}(a_i,a_j , a_v) \,,\nonumber\\[1ex]
&=&
\frac{3}{4} \frac{a^2_i a_v}{(a_i - a_v)^2} 
\left[ \frac{a^2_i \ln a_i }{(a_i - 1)^3} - \frac{a^2_v \ln a_v }{(a_v - 1)^3} \right] 
\nonumber\\[1ex]
&&
+ \,\,
\frac{a^2_i}{a_i - a_v} 
\left[ \frac{a_i  \ln a_i}{(a_i - 1)^3} - \frac{a_v \ln a_v }{(a_v - 1)^3} \right] 
+
\frac{a^3_i}{a_i - a_v} 
\left[ \frac{\ln a_i}{(a_i - 1)^3} - \frac{\ln a_v }{(a_v - 1)^3} \right] \nonumber\\[1ex]
&&
+ \frac{9}{4} \frac{a^5_i \ln a_i}{(a_i - a_v)(a_i - 1)^4}
- \frac{3}{4} \frac{a^4_i}{(a_i - a_v)(a_i - 1)^3} \nonumber\\[1ex]
&&
-\frac{a^2_i}{(a_i-1)^2 (a_v-1)} \cdot
\left[ 1-\frac{7}{4} a^2_i\right] 
\left[ \frac{2}{a_i-1}+ \frac{1}{a_v-1}\right] \nonumber\\[1ex]
&&
- \frac{a^2_i}{(a_i-1)^2(a_v-1)} \left[ \frac{3}{2} + \frac{1}{8} a^2_i \right]
\,,
\label{S1-VD}
\eeq

Some of the formulae simplify considerably in the limit 
\begin{equation} 
a_v \gg a_c,a_t \ , 
\end{equation} 
yielding: 
\beq
\Delta \bar{E}(a_c,a_t,a_v) &\simeq& 
-\frac{a_c}{a_v} 
\times \left\{ \frac{a^2_c \ln a_c}{(1-a_c)^3} -
\frac{1}{1-a_c} \left[ \frac{1}{a_t-1} -\frac{1}{1- a_c}+ \frac{3}{2}\right] \right\}\,,\\[1ex]
\Delta \bar{E}(a_i,a_v) &\simeq&-\frac{a^2_i}{a_v} \times 
\left[ \frac{\frac{3}{2}a^3_i+\frac{11}{4}a^2_i-2a_i }{(a_i-1)^4} \ln a_i +
\frac{\frac{1}{2} + \frac{3}{2}a_i - \frac{35}{8}a^2_i + \frac{1}{8}a^3_i}{(a_i-1)^2} \right]\,.
\eeq

\section{Wolfenstein's parametrization of the CKM matrix}
The Wolfenstein parameterization~\cite{Wolfenstein:1983yz} of the CKM matrix is:
\beq
V =
\bpm
\CKMud & \CKMus & \CKMub \\[2ex]
\CKMcd & \CKMcs & \CKMcb \\[2ex]
\CKMtd  & \CKMts & \CKMtb
\epm
+ {\cal O}(\lambda^6)\,,
\eeq
where $\lambda,A,\bar{\rho}=\rho (1 - \lambda^2/2),\bar{\eta}=\eta(1 - \lambda^2/2)$ are~\cite{Amsler:2008zzb}
\beq
\lambda = 0.2257^{+0.009}_{-0.001} \quad , \quad
A              = 0.814^{+0.021}_{-0.022} \quad , \quad 
\bar{\rho}= 0.135^{+0.031}_{-0.016} \quad ,\quad 
\bar{\eta}= 0.349^{+0.015}_{-0.017} \,,
\label{CKM-Wfs-parameter-PDG}
\eeq

\section{Correlation between $ |\epsilon_K|_{\rm SM}$ and $(\bar{\rho},\bar{\eta})$}

In our analysis, we have used the values of $\bar{\rho},\bar{\eta}$ in ~\cite{Amsler:2008zzb}.  It is instructive, however, to show how   $|\epsilon_K|_{\rm SM}$ modifies when the values  ($\bar{\rho},\bar{\eta}$) change by a small amount.  We indicate with $\left[ |\epsilon_K|_{\rm SM}\right]_{\rm new}$ the modified expression and with $\left[ |\epsilon_K|_{\rm SM}\right]_{\rm old}$ the initial value. We have then:
\beq
\left[ |\epsilon_K|_{\rm SM}\right]_{\rm new}
=
\left[ |\epsilon_K|_{\rm SM}\right]_{\rm old} \times \left( 1 + \frac{\Delta_{\bar{\eta}}}{\bar{\eta}_{\rm old}}\right)
-
C_{\cal E} \times \left( A^4\lambda^{10} \bar{\eta}\right)_{\rm old} \times \Delta_{\bar{\rho}}\,,
\eeq
where $\Delta_{\bar{\rho}(\bar{\eta})} \equiv \bar{\rho}(\bar{\eta})_{\rm new} - \bar{\rho}(\bar{\eta})_{\rm old}$ and 
\beq
C_{\cal E} \equiv \frac{G^2_F M^2_W}{12 \sqrt{2} \pi^2} \times \left[ \frac{M_K}{\Delta M_K} \right]_{\rm exp.} \hspace*{-3ex}
\times B_K f^2_K \times 2 \bar{E}_0(a_t) 
= 5.93 \times 10^4 \ .
\eeq
For example if we assume for $(\bar{\rho},\bar{\eta})_{\rm new} $ the values $ (0.136\pm0.032,0.340\pm0.016)$  \footnote{We thank Maurizio Pierini for stressing that these  are the best measured values for these quantities} one has
$\left[ |\epsilon_K|_{\rm SM}\right]_{\rm new} =  \left( 2.03^{+ 0.14}_{-0.12} \right) \times 10^{-3}$. This leads to  $\delta_{\epsilon}$ is $\delta_{\epsilon}= 0.100^{+ 0.076}_{-0.077} \,(68\%\cl)$.
This shows that at the $68 \%\cl$ we have always $\delta_{\epsilon} >0$. 
However in models of dynamical electroweak symmetry breaking we analyzed here  $\delta^{(\rm iVL)}_{\epsilon}$ is always negative since $W$ is always positive. Hence these value strongly reduce the allowed space of parameters  $(M_A,\tilde{g})$. We need to go to two sigmas to allow for the introduction of vector states.


\end{document}